\documentstyle[12pt]{article}

\begin{document}

\title{The infinite energy limit of the fine structure constant  equal to $1/4\pi$?}

\author{Antonio F. Ra\~nada\\Departamento de F\'{\i}sica Te\'orica\\Universidad Complutense, 28040 Madrid, Spain\thanks{E-mail: afr1@fis.ucm.es}}

\date{20 March 1999}

\maketitle

\begin{abstract}

A recently proposed topological mechanism for the quantization of the  charge gives the value $e_0=\sqrt{\hbar c}$ for both the fundamental electric and magnetic charges.  It is argued here that the corresponding fine structure constant $\alpha _0=1/4\pi$ could be interpreted as its value at infinite energy.

\end{abstract}

 This letter proposes an argument in favour of the idea that the infinite energy limit of the fine structure constant is equal to $1/4\pi$. The argument is based on two grounds: (i) a recent topological mechanism for charge quantization which implies that the fundamental electric and magnetic charges are both equal to $e_0=\sqrt{\hbar c}=3.3\,e$, the corresponding fine structure constant being $\alpha _0=1/4\pi$ \cite{Ran98}; and (ii) the appealing and plausible idea that, in the limit of very high energies, the interactions of charged particles could be determined by their bare charges (this meaning the value that their charges would have if they were not renormalized by the quantum vacuum, see for instance section 11.8 of \cite{Mil94}). A warning is however necessary: the concept of bare charge is more complex than what was thought some time ago, so that it is better now to speak of charge at a certain scale. To be precise,  when the expression ``bare charge" will be used here, it will be taken as equivalent and  synonymous to ``infinite energy limit of the charge" or, more correctly, ``charge at infinite momentum transfer", defined as  $e_\infty =\sqrt{4\pi \hbar c\alpha _\infty}$, where $\alpha _\infty =\lim \alpha (Q^2)$ when $Q^2\rightarrow \infty$.

   The possibility of a finite value for $\alpha _\infty$ is an intriguing idea worth of study.  Indeed, it was discussed very early by Gell-Mann and Low  in their classic and seminal paper ``QED at small distances" \cite{Gel54}, in which they showed that it is something to be seriously considered. However, they could not decide with their analysis whether $e_\infty$ is finite or infinite. The standard QED statement that it is infinite was established later because of perturbative calculations, but it can not be said that the alternative presented by Gell-Mann and Low had been definitely settled.

The infinite energy charge $e_\infty$  of an electron is partially screened by the sea of virtual pairs that are continuously being created and destroyed in empty space. It is hence said that it is renormalized. As the pairs are polarized, they generate a cloud of polarization charge near any charged particle, with the result that the observed value of the charge is smaller than $e_\infty$. Moreover, the apparent electron charge increases as any probe goes deeper into the polarization cloud and is therefore less screened. This effect is difficult to measure,  as it can only be appreciated at extremely short distances, but it has been observed indeed in experiments of electron-positron scattering at high energies \cite{Lev97}. In other words: the vacuum is dielectric. On the other hand, it is paramagnetic, since its effect on the magnetic field is due to the spin of the pairs. As a consequence, the hypothetical magnetic charge would be observed with a greater value at low energy  than at very high energy, contrariwise to the electron charge.

The name bare charge is appropriate for $e_\infty$, as it is easy to understand intuitively. When two electrons interact with very high momentum transfer, each one is so deeply inside the polarization cloud around the other that no space is left between them to screen their charges, so that the bare values, i.e. $e_\infty$, interact directly. As unification is assumed to occur at very high energy, it is an appealing idea that $\alpha _\infty =\alpha _{\rm GUT}$ (it is true that one could imagine that $\alpha (Q^2)$ has a plateau at the unification scale corresponding to a critical value smaller than $\alpha _\infty$, but we assume the simpler situation in which that plateau does not exist). This suggests that a unified theory could be a theory of bare particles (in the sense of neglecting the effect of the vacuum). If this were the case, nature would have provided us with a natural cutoff, in such a way that  $\alpha _{\rm GUT}= \alpha _\infty$.

The charge quantization mechanism given in \cite{Ran98} is based on a topological model endowed with a  structure induced by the topology of the magnetic and electric force lines, which are represented as the level curves of a couple of complex scalar fields $\phi ,\theta$ \cite{Ran98,Ran92,Ran95,Ran97}, the electromagnetic tensor $F_{\mu\nu}$ being expressed in terms of these scalars by a certain precise transformation $T:\phi ,\theta \mapsto F_{\mu\nu}$.
 The scalars $\phi, \theta$ obey highly nonlinear equations. Surprisingly however these nonlinear equations are transformed exactly into Maxwell equations by the transformation $T$. Consequently,  the $F_{\mu\nu}$ of the model are standard Maxwell fields (although behaving in a particular way around the infinity), so that it
is equivalent to Maxwell standard theory in any bounded spacetime domain.

A consequence of that topological structure is that the charge inside any volume is always equal to $n\sqrt{\hbar c}$, the integer $n$ being understood as the degree of a map between two spheres. It turns out that each electric line around a point charge is labelled by a complex number, the value of $\theta ({\bf r},t)$ along it, in such a way that there are exactly $n$ lines with the same label, taking into account the orientation of the map (the same would apply to a magnetic charge, with $\phi$ instead of $\theta$). 
As this topological mechanism operates at the classical level and since the charge is necessarily affected by the quantum vacuum \cite{Mil94} to give the dressed observed value, the fundamental charge $e_0=\sqrt{\hbar c}$ must be interpreted as the infinite energy value of both the electric and magnetic charges $e_\infty$ and $g_\infty$.  In other words, the model predicts that $e_\infty =g_\infty =e_0$.

(It is perhaps worth mentioning that, in a different context, these topological ideas have inspired a model of ball lightning in which this phenomenon is assumed be a magnetic knot coupled to a plasma \cite{Ran96,Ran98b}. The linking of magnetic lines turns out to have a stabilizing effect which allow the fireballs to last for much more time than expected.)

As a consequence of these considerations, the argument announced in the first line which leads to the equalities $\alpha _{\rm GUT}=\alpha _\infty =1/4\pi$ goes as follows:

1. The value of the fundamental charge implied by the topological mechanism \cite{Ran98} $e_0=\sqrt{\hbar c}$ is in the right interval to verify $e_0= e_\infty =g_\infty$, that is to be equal to the common value of both the fundamental electric and magnetic infinite energy charges. This is so because, as the quantum vacuum is dielectric but paramagnetic, the following inequality must be satisfied then: $e<e_0 <g$, as it is indeed, since $e=0.3028$, $e_0=1$, $g=e/2\alpha = 20.75$, in natural units.
 
Note that it is impossible to have a complete symmetry between electricity and magnetism simultaneously at low and high energy. The lack of symmetry between the electron and the Dirac monopole charges would be due, in this view, to the vacuum polarization: according to the topological model, the electric and magnetic infinite energy charges are equal and verify $e_\infty g_\infty=e_0^2=1$, but they would be decreased and increased, respectively,  by the sea of virtual pairs, until the electron and the monopole charge values verifying the Dirac relation $eg=2\pi$ \cite{Dir31}. The qualitative picture seems nice and appealing.

2. Let us admit as a working hypothesis that two charged particles interact with their bare charges in the limit of very high energies (as explained above). There could be then a conflict between (i) a unified theory of electroweak and strong forces, in which $\alpha =\alpha _s$ at very high energies, and (ii) an infinite value of $\alpha _\infty$. This is so because unification implies that the curves of the running constants $\alpha (Q^2)$ and $\alpha _s(Q^2)$ must converge asymptotically to the same value $\alpha _{\rm GUT}$. It could be argued that, to have unification at a certain scale, it would be enough that these two curves be close in an energy interval, even if they cross and separate afterwards. But, in that case, the unified theory would  be just an approximate accident at certain energy interval. On the other hand, the assumption that both running constants go asymptotically to the same finite value $\alpha _{\rm GUT}$ gives a much deeper meaning to the idea of unified theory, and is therefore much more appealing. In that case,  $e_\infty$ must be expected to be finite, and the equality $\alpha _{\rm GUT}=\alpha _\infty$ must be satisfied.

3. The value $\alpha _0=e_0^2/4\pi \hbar c =1/4\pi =0.0796$ for the infinite energy fine structure constant $\alpha _\infty$ is thought provoking and fitting, since $\alpha _{\rm GUT}$ is believed to be in the interval $(0.05,\,0.1)$ (some say furthermore that close to 0.08). This reaffirms the assert that the fundamental value of the charge given by the topological mechanism $e_0$ could be equal to $e_\infty$, the infinite energy electron charge (and the infinite energy monopole charge also), and supports the statement that $\alpha _{\rm GUT}$ must be equal to $\alpha _0$ and to $1/4\pi$.
 All this is certainly curious and intriguing since the topological mechanism for the quantization of the charge \cite{Ran98} is obtained just by putting some topology in elementary classical low energy electrodynamics \cite{Ran92}.

The conclusion of this letter is that the following three ideas must be studied carefully: (i) the complete symmetry between electricity and magnetism at the level of the infinite energy charges, both being equal to $\sqrt{\hbar c}$, the symmetry being broken by the dielectric and paramagnetic quantum vacuum; (ii) that the topological model on which the topological mechanism of quantization is based could give a theory of high energy electromagentism at the unification scale; and (iii) that the value which it predicts for the infinite energy fine structure constant $\alpha _0=1/4\pi$ could be equal to $\alpha _\infty$ and also to $\alpha _{\rm GUT}$, the constant of the unified theory of strong and electroweak interactions. In this way the three quantities, both the electric and magnetic fine structure constants at infinite momnetum transfer and $\alpha _{\rm GUT}$, would be equal and there would be a complete symmetry between electricity, magnetism and strong force at the level of bare particles (i.e. at $Q^2=\infty$), this symmetry being broken by the effect of the quantum vacuum.

\end{document}